# Femtosecond Laser Crystallization of Ultrathin a-Ge Films in Multilayer Stacks with Silicon Layers

Yuzhu Cheng [1], Alexander V. Bulgakov [2], Nadezhda M. Bulgakova [2,*], Jiří Beránek [2,3], Aleksey V. Kacyuba [4], and Vladimir A. Volodin [1,4]

[1] Physics Department, Novosibirsk State University, Pirogova Street, 2, Novosibirsk, 630090 Russia; chengyuzhu9@gmail.com (Y.C.); kacyuba@isp.nsc.ru (A.V.K.); volodin@isp.nsc.ru (V.A.V.)
[2] FZU - Institute of Physics of the Czech Academy of Sciences, Na Slovance 1999/2, 182 00 Praha 8, Czech Republic; nadezhda.bulgakova@fzu.cz (N.M.B.); alexander.bulgakov@fzu.cz (A.V.B.); jiri.beranek@fzu.cz (J.B.)
[3] Faculty of Nuclear Sciences and Physical Engineering, Czech Technical University in Prague, Trojanova 13, 12001 Prague, Czech Republic
[4] Rzhanov Institute of Semiconductor Physics, Siberian Branch, Russian Academy of Sciences, Lavrentiev Ave, 13, Novosibirsk, 630090 Russia; volodin@isp.nsc.ru; kacyuba@isp.nsc.ru (A.V.K.)

* Correspondence: nadezhda.bulgakova@fzu.cz

**Abstract**

Ultrashort pulsed laser annealing is an efficient technique for crystallizing amorphous semiconductors with the possibility to obtain polycrystalline films at low temperatures, below the melting point, through non-thermal processes. Here, a multilayer structure consisting of alternating amorphous silicon and germanium films was annealed by mid-infrared (1500 nm) ultrashort (70 fs) laser pulses under single-shot and multi-shot irradiation conditions. We investigate selective crystallization of ultrathin (3.5 nm) a-Ge film, promising for the generation of highly photostable nanodots. Based on Raman spectroscopy analysis, we demonstrate that, in contrast to thicker (above 10 nm) Ge films, explosive stress-induced crystallization is suppressed in such ultrathin systems and proceeds via thermal melting. This is likely due to the islet structure of ultrathin films which results in the formation of nanopores at the Si-Ge interface and reduces stress confinement during ultrashort laser heating.

## 1. Introduction

Semiconductor nanomaterials are one of the blocks for future technologies due to their unique optical, electrical, thermal, and mechanical properties superior over their bulk counterparts [1]. However, upon semiconductor nanoparticle synthesis and thin film fabrication, the as-prepared nanomaterial is usually amorphous while, for their use in real applications such as sensors, solar energy harvesting, or thin-film transistors, crystalline nanosized materials are of high demand [2,3]. One of the methods of preparing nanocrystals and nanocrystalline thin films is annealing of amorphous materials [4]. The traditional method is thermal annealing of the whole structure in a furnace [5] that does not allow to perform selective crystallization of a certain part of the film or a specific film component. These limitations can be overcome by using laser annealing [6-9]. A variety of available lasers enables to choose laser wavelengths and pulse durations to achieve optimal conditions for selective crystallization [7,9].

One of the advantages of short and ultrashort pulse laser annealing is that, due to localized rapid laser heating, the heat-affected zone is very small on the scale of the entire sample, and the absorbed laser energy rapidly dissipates [10]. This allows for controlling the depth of crystallization of amorphous material [11] and can be used for gently crystallizing refractory amorphous films on non-refractory substrates without thermal damage to the latter [9]. Furthermore, crystallization of semiconductors with ultrashort laser pulses can be achieved, under certain conditions, at fairly low temperatures, below the material melting point [8]. Using pulsed laser annealing, it is possible to produce thin-film transistors (TFTs) based on crystallization of a-Si:H at relatively low levels of heating. Already first attempts to fabricate TFTs using pulsed laser annealing of amorphous silicon have shown good characteristics with carrier mobility of ~0.6 cm$^2$/V·s [12]. In further works, the approaches were developed for lateral crystallization. It was shown that, after laser annealing of amorphous silicon, it was possible to achieve carrier mobility up to 410 cm$^2$/V·s [13].

The mentioned features of pulsed laser annealing that enable utilizing non-refractory, relatively cheap, and flexible substrates is highly beneficial for the fabrication of solar cells, as for their mass production, there is a need for cheap substrate materials. In recent years, much effort has also been directed to improving solar cell elements by including nanostructures of metals and narrow-gap semiconductors in their matrix, which results in enhanced light absorption [14-15]. In particular, germanium-based nanostructures and nanofilms attract much attention for applications in the fields of flexible electronics and thin-film solar elements with improved characteristics [16,17]. Ge/Si heterostructures as well as multilayer nanostructures, which consist of germanium inclusions in a wider-bandgap silicon, are of both fundamental and practical interest [18]. Heterostructures with germanium quantum dots in a silicon matrix integrated into resonator devices allow reaching quantum efficiency sufficient for the use in silicon optoelectronics [19]. Incorporating germanium and tin nanoparticles into unalloyed layers of p-i-n structures based on amorphous or microcrystalline silicon films enables to enhance the efficiency of solar cells [20] and to widen the absorption spectrum toward long wavelengths for solar elements and photodiodes [21]. It is known that the difference between the lattice constants of germanium and silicon plays a significant role in the mechanisms of self-organization of nanostructures (quantum dots) of germanium [22]. Modulation-doped field-effect transistors Si/SiGe have demonstrated potential for high-speed applications [23].

Therefore, Ge/Si heterostructures based on amorphous materials can be fabricated on large areas and on non-refractory substrates, and, thus, it is important to develop methods for their crystallization by laser irradiation. In this paper, we present the results of an attempt to crystallize ultrathin (3.5 nm) germanium layers in multilayer stacks with silicon films to explore a possibility of formation of periodically located crystallized Ge nanolayers and/or Ge quantum dots which would be suitable for applications in optoelectronics. In sections 2, the preparation of Si/Ge multilayers stacks is described as well as the choice of irradiation conditions and the sample analysis have been provided. Section 3 presents a detailed description of the results and, in section 4, we discuss why the laser crystallization process in multilayer stack containing few-nm Ge films behaves differently as compared to thicker Ge films as in [9]. Finally, in section 5, conclusions are made with outlining further prospectives for Ge/Si multilayer structures and their applications.

## 2. Materials and Methods

The Si/Ge multilayer stacks (MLS) consisting of alternating amorphous Si and Ge nanolayers with thicknesses of 30 and 3.5 nm, respectively, with a Si layer on the top, were fabricated by molecular-beam epitaxy (MBE) on a 1-mm-thick glass substrate. During the deposition, the temperature of the glass substrate remained near the room temperature, and the deposition rate was 1 Å/s for Si and 0.1 Å/s for Ge. The thickness of the layers was controlled by the deposition rate and deposition time. The grown MLS structures did not contain hydrogen, contrary to the multilayer stacks studied in [9]. The schematics of MLS and its irradiation are shown in Figure 1.

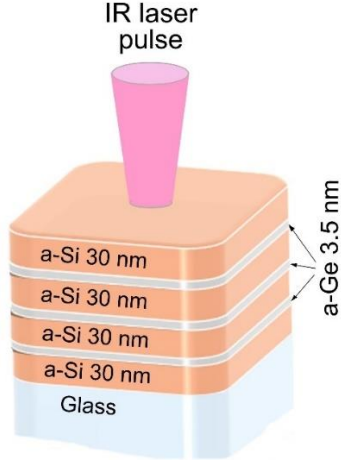

**Figure 1.** Schematics of a-Ge/a-Si multilayer stack irradiated by laser.

Modification of the Ge/Si stacks was performed using a femtosecond Ti:sapphire laser Astrella (Coherent, Santa Clara, CA, USA) equipped with an optical parametric amplifier (TOPAS from Light Conversion, Vilnius, Lithuania). For the modification experiments, 70-fs Gaussian pulses in the mid-IR spectral range (wavelength 1500 nm) with the pulse energy up to 400 µJ were chosen. These irradiation parameters were found previously as favorable ones for laser annealing of amorphous semiconductors [9,24]. The laser beam was focused at normal incidence on the sample surface by a glass lens (focal length 150 mm) onto a circular spot with an effective diameter of $2w_0 = 0.53$ µm. The spot size was determined in separate calibration experiments using the $D^2$ technique [25] under single-shot conditions. The pulse energy $E_0$ was varied by an attenuator consisting of a $\lambda/2$ plate and a polarizer to obtain the peak fluence $F_0 = 2E_0/\pi w_0^2$ in the range of 20-200 mJ/cm². Details of the annealing experiments can be found in [9,24].

Phase composition of the irradiated structure was investigated using a micro-Raman (spectrometer T64000 from Horiba Jobin Yvon, Lille, France), which utilizes a fiber laser at a wavelength of 514.5 nm. The laser beam was focused to a spot of 10 µm diameter. The Raman spectra were collected from the regions within the center of the irradiation spot of the annealing laser, which can be considered as uniformly annealed.

Before the laser annealing, the transmission and reflection of the Ge/Si stacks were measured in the range of wavelengths from 500 nm to 2500 nm using a spectrophotometer UV-3600 (Shimadzu, Kyoto, Japan). To register the reflection spectra, a specular reflection attachment with an angle of incidence of 5 degrees was used. As a reference, an aluminum mirror was used, and a correction was made taking into account the Al reflectivity. It was not possible to study the transmission and reflection spectra after laser annealing because the uniformly annealed spots were too small.

## 3. Results

The transmission and reflection spectra of an as-deposited a-Ge/a-Si stack are shown in Figure 2. The minima and maxima seen in the spectra correspond to the interference inside the structure. The thickness of the total stack estimated from the interference is ~130 nm that is in agreement with that obtained from the growth rates of germanium and silicon nanolayers. It is necessary to mention that the sum of the reflectance and transmittance is only slightly lower than 100%. The absorption coefficient of the non-hydrogenated amorphous silicon for the photon energy of 0.827 eV is ~5×10³ cm⁻¹ [26]. For the total thickness of a-Ge layers of ~10 nm, the linear absorption can be estimated as ~0.5% while non-linear effects are negligible at the laser intensities used for measurements [9]. The structure of the irradiation spots produced on the Si/Ge MLS is similar to that observed in [9] and, at fairly high laser fluences, consists of three zones: external modification region, middle damage region, and central ablation region. The threshold fluences for the appearance of these regions were measured using the $D^2$ method for Gaussian pulses [27]. The obtained threshold values for modification, damage, and ablation are only slightly lower or equal to those of hydrogenated a-Ge/a-Si layers [9] and are respectively 45, 70, and 110 mJ/cm$^2$. We note that, in the case of ultrathin Ge layers, the threshold fluences are expected to be governed by silicon, which, being the top layer of the studied multilayer stack, is a first shield experiencing laser radiation impact (Figure 1). It is known that the linear absorption coefficient of non-hydrogenated silicon is higher than that of hydrogenated ones [28] and the same tendency should be expected for two-photon absorption. However, the reflectivity of non-hydrogenated a-Si is also higher [28] that likely serves as a compensating factor resulting in the similar modification, damage, and ablation thresholds.

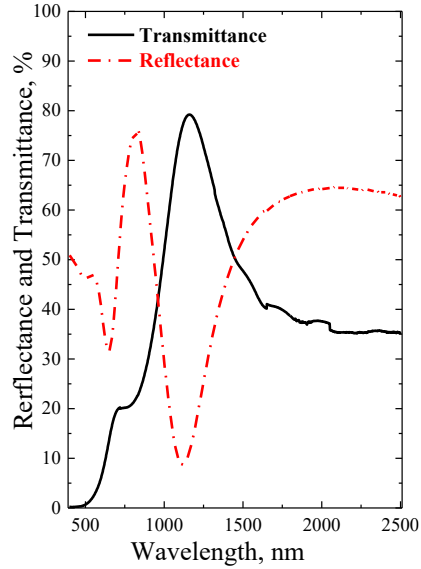

**Figure 2.** Transmittance and reflectance spectra of an as-deposited a-Si/a-Ge multilayer stack on a glass substrate.

The Raman spectrum of the as-deposited a-Ge/a-Si MLS is presented in Figure 3. The spectrum contains wide bands at 150, 280, and 480 cm$^{-1}$. These bands correspond to the maxima of the densities of vibrational states for amorphous silicon and germanium. The maxima at 150 and 480 cm$^{-1}$ are conditioned by light scattering on the local vibrations of Si-Si bonds of acoustic (neighboring atoms oscillate in phase) and optical (antiphase oscillations) types [29]. The maximum at 280 cm$^{-1}$ corresponds to light scattering on the vibrational states of Ge-Ge bonds of the optical type [30]. The maximum in the densities of vibrational states of the acoustic type for germanium is located at ~100 cm$^{-1}$ [30] which cannot be recognized in the Raman spectrum as this range is cut off by the edge filter. Narrow lines seen at frequencies smaller than 150 cm$^{-1}$ are attributable to scattering on vibrational-rotational modes of the molecules in air [31]. It is known that the vibrational optical frequency of Ge-Si bonds is approximately 400 cm$^{-1}$ [32], which is shifting with changing Ge fraction [33]. It is difficult to recognize the presence of these bonds in the Raman spectra against the background of a wide asymmetric band of Si-Si bonds. Thus, the as-deposited samples consist of amorphous germanium and amorphous silicon.

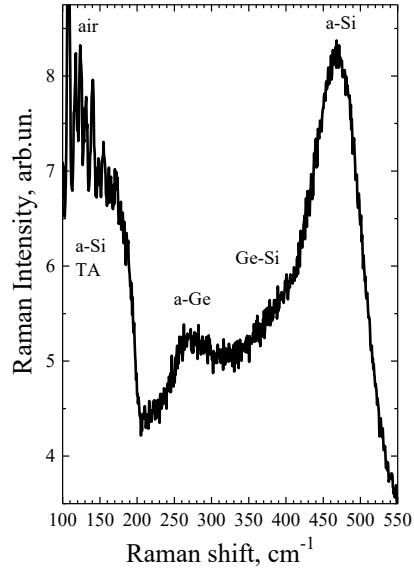

**Figure 3.** Raman spectrum of as-deposited Si/Ge MLS on the glass substrate.

In Figure 4a, the Raman spectra are shown for the Ge/Si MLS after irradiation by femtosecond laser pulses at relatively high laser fluences of 155 and 130 mJ/cm$^2$. The number of laser shots onto one spot was 10. The spectra correspond generally to the Ge-Si solid solutions. The vibrational frequencies of Ge-Si bonds lie in the range from ~390 cm$^{-1}$ to 440 cm$^{-1}$ [32]. From the spectra in Figure 4, a conclusion can be made that the peaks in this range appear from the oscillations of the Ge-Si bonds, thus indicating that intermixing of the layers of germanium and silicon occurs with subsequent crystallization.

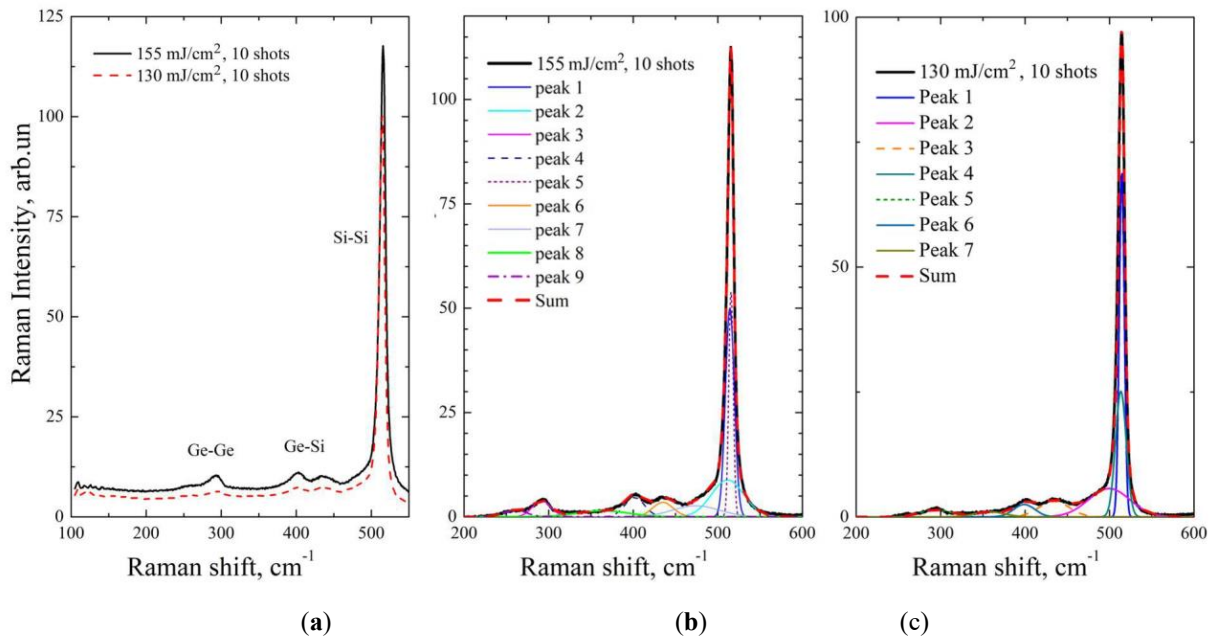

**Figure 4.** (a) Raman spectra of Si/Ge MLS on glass substrate after femtosecond laser annealing with 10 pulses per irradiation spot at high laser fluences. (b), (c) Deconvolution of the Raman spectra into individual Gaussian peaks at fluences of 155 mJ/cm$^2$ and 130 mJ/cm$^2$ respectively.

The compositions of the solid solutions of $Ge_xSi_{(1-x)}$ after laser irradiation with the high and medium fluences were determined using the following method. If the solution is completely intermixed, the concentrations of bonds Ge-Ge, Si-Si, and Ge-Si depend on the stoichiometric factor $x$ as [32,34-36]

$$N_{\text{Ge-Ge}} = x^2; \quad N_{\text{Ge-Si}} = 2x(1-x); \quad N_{\text{Si-Si}} = (1-x)^2.$$

For determining the composition, the following expressions were used

$$\frac{I_{\text{Si-Si}}}{I_{\text{Ge-Si}}} = A\frac{(1-x)}{2x}, \tag{1}$$

$$\frac{I_{\text{Ge-Ge}}}{I_{\text{Ge-Si}}} = B\frac{x}{2(1-x)}. \tag{2}$$

The values $I_{\text{Si-Si}}$, $I_{\text{Ge-Si}}$, and $I_{\text{Ge-Ge}}$ are the integral intensities of the Raman spectra for Si-Si, Ge-Si, and Ge-Ge bonds, respectively. The coefficients $A$ and $B$, which are determined by the composition-dependent cross-sections of the Raman spectra for different bond types, were deduced in [32].

To analyze the Ge-Si intermixing efficiency, the Raman spectra were deconvoluted into Gaussian curves (Figures 4b and 4c) using the peak fitting software Fityk [37]. For annealing with 10 laser pulses at 155 mJ/cm², the $x$ factor is equal to 0.13 and 0.34 according to equations (1) and (2), respectively. This indicates that not all layers are mixed uniformly, and there are some regions enriched by germanium and extended regions of pure silicon. The integral intensity $I_{\text{Si-Si}}$ in equation (1) contains signals from the regions where silicon is non-mixed and mixed with germanium. We note that according to both equations, the content of germanium in the solid solution is higher than it should be based on the ratio of the thicknesses of the layers of germanium and silicon (uniform intermixing assumes the composition of approximately $Ge_{0.08}Si_{0.92}$). The observed Ge-excessive composition indicates that the silicon layers contain both germanium nanocrystals and the regions of crystalline germanium, which are separated by transition layers of solid solution.

After irradiation of Ge/Si MLS with 10 laser pulses at a laser fluence of 130 mJ/cm² (Figure 4a and 4c), the bands corresponding to Ge-Ge bonds can still be recognized. This allows estimating the composition of solid solution as $x = 0.1$ and $x = 0.23$ from equations (1) and (2), respectively, which indicates that the solid solution is more homogeneous. We note that unmixed silicon can partially be in an amorphous state, which can explain the estimated enhanced concentration of germanium. Indeed, the crystalline Si-Si peaks have shoulders toward the amorphous phase (compare with Figure 3) that point to the presence of an overlapping amorphous peak. At excitation of the Raman spectrum by green laser light (514.5 nm), the Raman cross-section is larger for crystalline silicon as compared with the amorphous one [38]. As a result, the equations (1) and (2) give underestimated content of silicon and, respectively, overestimated content of germanium.

At a laser fluence of 80 mJ/cm² (10 pulses, Figure 5), the composition is $Ge_{0.08}Si_{0.92}$ according to equations (1) and (2). This corresponds exactly to the composition estimated from the ratio of the thicknesses of the Ge and Si layers. We assume that, in this regime, a complete intermixing of layers takes place. Interestingly, this happens in the regimes of moderate fluences. It can be speculated that, in the case of large fluences (Figure 4), the complete melting of layers occurs. During cooling, the regions enriched by silicon first crystallize and nanoislets of germanium precipitate. However, note that, at single pulse irradiation at a fluence of 80 mJ/cm² (Figure 5), Ge-Ge peak is not recognized while Si crystalline peak starts to emerge.

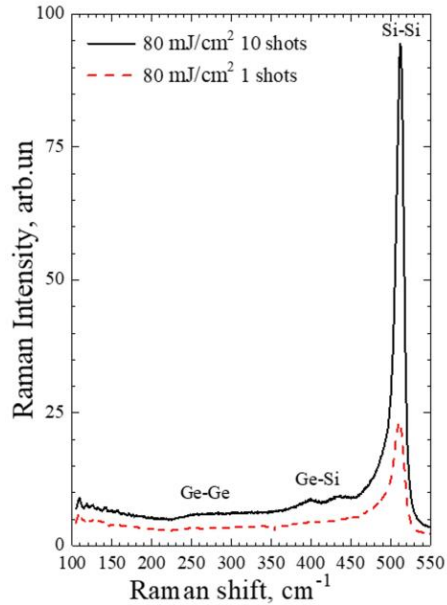

**Figure 5.** Raman spectra of Si/Ge MLS on glass substrate after laser annealing with 1 and 10 pulses per irradiation spot at a medium fluence of 80 mJ/cm$^2$.

Presumably, in the case of formation of the solid solution Ge$_{0.08}$Si$_{0.92}$ (Figure 5), the complete melting is not reached, and interdiffusion of germanium and silicon can proceed partially in the solid state. At one laser shot, the complete intermixing does not happen. The germanium peak is not seen, which indicates that germanium remains in the amorphous state. It is known that, under Raman spectrum excitation with green light (514.5 nm), the Raman cross-section for crystalline germanium is considerably larger than for amorphous germanium [39]. If the crystalline phase is present, there should be a clear manifestation of the Ge crystalline peak. The broadened Si peak indicates that silicon is not completely crystallized.

At a lower laser fluence of 66 mJ/cm$^2$ (Figure 6, left, 10 pulses), it looks like germanium remains in the amorphous phase. However, due to ultrathin layers of germanium, the most Ge fraction is intermixed with silicon, creating a pronounced Ge-Si peak, while crystalline germanium phase, if present, is not well distinguished. The silicon crystalline Si-Si peak is very pronounced although the shoulder toward the amorphous phase is even more extended as compared with Figures 4 and 5.

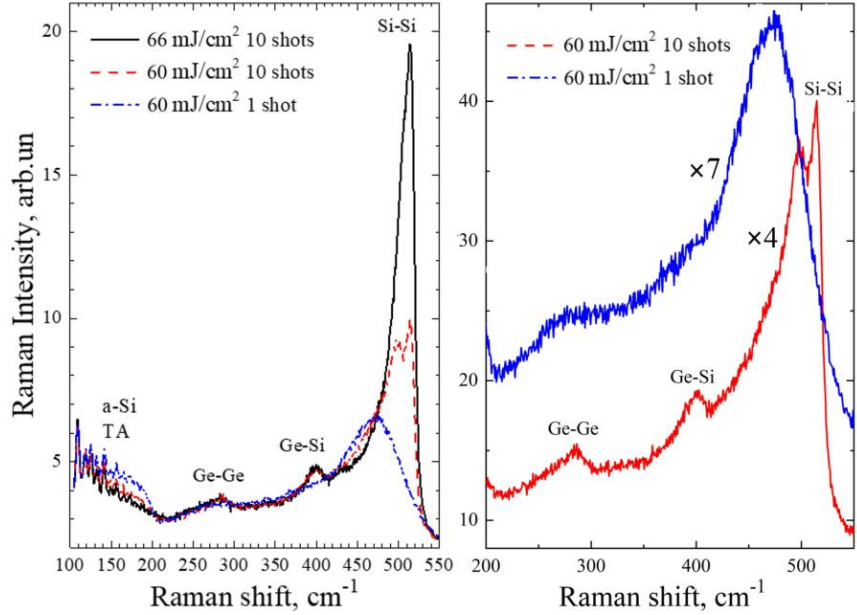

**Figure 6.** Left: Raman spectra of Si/Ge MLS on glass substrate after laser annealing at low fluence. Right: Magnified view of the Raman spectra obtained for 60 mJ/cm2 at 1 shot (blue line, ×7) and 10 shots (red line, ×4).

With further decreasing laser fluence (Figure 6, 60 mJ/cm$^2$, left, 10 pulses), a considerable part of silicon remains unmixed as a peak of nanocrystalline silicon at 513 cm$^{-1}$ is present. Additionally, a peak with location at 495 cm$^{-1}$ is pronounced, which can be attributed to a solid Ge-Si solution. Indeed, if to apply the experimental fits for the bond frequencies obtained for Ge-Si solution in Ref. [40]

$$\omega_{\text{Ge-Ge}} = 300.3 - 32(1-x) + 12(1-x)^2,$$
$$\omega_{\text{Si-Si}} = 520 - 62x, \qquad (3)$$

this peak should correspond to Si-Si oscillations in the solution Ge$_{0.4}$Si$_{0.6}$. Apparently, germanium has been mixed with nearest atomic layers of silicon with formation of a solid solution. The remaining silicon is partially crystallized as evidenced by the peak with position at 513 cm$^{-3}$. According to estimations as in [32], the size of crystallites is about 3 nm. For single pulse irradiation at 60 mJ/cm$^2$ that is below the modification threshold, the Raman spectrum is very similar to that of as prepared MLS (compare with Figure 3). We note that namely at such irradiation conditions, at fluences slightly below the modification threshold, formation of Ge crystallites without signs of mixing between germanium and silicon in the MLS was achieved in [9]. The possible reasons for difference between the results of work [9] and the present results for ultrathin Ge layers will be discussed in the next section.

## 4. Discussion

One of the initial goals of this study was to achieve crystallization of ultrathin (few-nm) germanium films in amorphous non-hydrogenated Ge/Si multilayer stacks with preserving the amorphous phase of silicon and avoiding intermixing between adjacent layers, similarly to what was achieved for Ge/Si MLS with thicker Ge layers [9]. However, as seen from the above results, such a regime is not detected under our irradiation conditions. In this section we discuss the difference of the physical properties and processes, which did not allow us here to observe a selective crystallization of germanium in Ge/Si multilayer stacks in comparison with the thicker Ge layers as in [9].

In work [9], successful Ge crystallization without intermixing with the adjacent Si layers was found at single pulses with laser fluence slightly below the modification threshold while no crystallization signs have been detected in the present experiments at such laser fluences. When applying several pulses, up to 10 shots, a partial crystallization of silicon occurs while Ge-Si intermixing is evident with the formation of solid Ge-Si solution. Figure 6, right, demonstrates magnified Raman spectra for laser-irradiated MLS at a laser fluence of 60 mJ/cm$^2$ with 1 and 10 pulses presented on the left of Figure 6. If single-pulse-irradiated stacks do not show any signs of crystallization (compare with Figure 3 for as-deposited MLS), a small peak of crystalline germanium can be recognized at 10 laser pulses. The peak maximum is at ~287 cm$^{-1}$, red shifted as compared to that of ~4.5-nm crystallites located at 298.5 cm$^{-1}$ observed in [9], indicating formation of tiny Ge crystallites (quantum dot) whose average size is to be determined. Indeed, the red shift of Raman spectra is increasing with decreasing nanocrystal size with simultaneous peak broadening [41,42] that is in agreement with the Ge peak in Figure 6, right. According to the Zi's model [42,43], it can be expected that the average size of Ge crystallites is 1-1.5 nm.

Now the question arises why Ge nanolayers do not crystallize in the "cold" manner as was proposed in [8]. The main reason can be attributed to the ultrathin layers of germanium. When amorphous germanium layers are deposited by molecular-beam epitaxy, they first form islets which, with increasing the deposited material, transfer to continuous layers [43]. We anticipate that, at ultrathin Ge films are deposited as 0D nanostructures, which, being covered by thicker silicon layers, unavoidably form nanopores. Such sandwich structures with nanopores at interfaces should demonstrate changes in physical properties, e.g. in thermal diffusivity [44], as compared to their pore-free counterparts. In the case of laser annealing under our irradiation conditions, germanium nanolayers are readily absorbing laser irradiation, heat, and expand. For relatively thick Ge films like in [8,9], a-Ge layers are strongly confined by the adjacent silicon layers, thus experiencing strong stress with the possibility of explosive crystallization below the melting threshold. For 3.5-nm layers, the pores are expected to be of size comparable with film thickness and serve as the stress-release centers upon thermal expansion. As a result, the stress-confinement regime cannot be realized, and Ge crystallization proceeds via thermal melting. However, more studies are envisioned for the multilayer formation of Ge nanodots which can find application niches in optoelectronics, solar cells, and photodetectors.

## 5. Conclusions

In this work, an attempt has been made to selectively crystallize ultrathin (3.5 nm) amorphous non-hydrogenated germanium layers in the multilayer stacks with amorphous non-hydrogenated silicon films of 30-nm thickness by ultrashort mid-IR laser pulses at 1500 nm wavelength. Such pulses were successfully used for laser annealing of thicker a-Ge layers in a-Ge/a-Si stacks with formation of a-Ge nanocrytallites without intermixing with the adjacent silicon layers [8,9]. Selective crystallization of thinner non-hydrogenated Ge films can enable the formation of periodic nanodots of high photostability embedded in a silicon matrix, which can find applications in optoelectronics, photodetectors, solar elements, etc.

However, the laser annealing results have demonstrated a different character of crystallization as compared to those for thicker hydrogenated counterparts. At relatively low laser fluences, at which selective crystallization was achieved for thicker a-Ge:H layers by single laser pulses, no change in Raman spectra was observed for ultrathin a-Ge films as compared to as-prepared samples. Crystallization of a-Ge layers was found at multi-shot irradiation when intermixing of germanium and silicon is observed with formation of solid Ge-Si solution and partial crystallization of silicon. The evident absence of the explosive crystallization pathways revealed in [8,9] for thicker layers is explained by peculiarities of deposition of ultrathin films, plausibly providing nanopores at interfaces in multilayer structures. Nanopores reduce stress confinement for nanofilms thermally expanding under laser heating, thus eliminating conditions for non-thermal crystallization. This manuscript highlights novel, subtle aspects of laser processing of nanostructured material, which can be important for achieving desired laser-modification results.

**Funding:** A.V.B., N.M.B., and J.B. acknowledge support from the European Regional Development Fund and the State Budget of the Czech Republic (Project SENDISO No. CZ.02.01.01/00/22_008/0004596). The study of A.V.K. was supported by the Ministry of Science and Higher Education of the Russian Federation, (theme No. FWGW-2025-0023). The study of V.A.V. and Y.C. was supported by the Ministry of Science and Higher Education of the Russian Federation, (theme No. FSUS-2024-0020). The study of Y.C. was also supported by Program of China Scholarship Council, Grant No. 202310100100.

**Acknowledgments:** The authors are grateful to "VTAN" NSU for providing equipment for Raman spectroscopy and HRTEM image analysis.

**Conflicts of Interest:** The authors declare no conflicts of interest.

## References

1. Borca, B.; Bartha, C. Advances of nanoparticles and thin films. *Coatings* **2022**, *12*, 1138.
2. Sakaike, K.; Higashi S.; Murakami, H.; Miyazaki, S. Crystallization of amorphous Ge films induced by semiconductor diode laser annealing. *Thin Solid Films* **2008**, *516*, 3595-3600.
3. Imajo, T.; Ishiyama, T.; Saitoh, N.; Yoshizawa, N.; Suemasu, T.; Toko, K. Record-high hole mobility germanium on flexible plastic with controlled interfacial reaction. *ACS Appl. Electron. Mater*. **2022**, *4*, 269–275 (2022).
4. Rebohle, L.; Prucnal, S.; Skorupa, W. A review of thermal processing in the subsecond range: semiconductors and beyond. *Semicond. Sci. Technol*. **2016**, *31*, 103001.
5. Spinella, C.; Lombardo, S.; Priolo, F. Crystal grain nucleation in amorphous silicon. *J. Appl. Phys*. **1998**, *84*, 5383–5414.
6. Poate, J. M.; Mayer J. W. (Eds). *Laser Annealing of Semiconductors*; Academic Press: New York, NY, USA, 1982.
7. Aktas, O.; Peacock, A. C. Laser thermal processing of group IV semiconductors for integrated photonic systems. *Adv. Phot. Res*. **2021**, *2*, 2000159.
8. Mirza, I.; Bulgakov, A. V.; Sopha, H.; Starinskiy, S. V.; Turčičová, H.; Novák, O.; Mužík, J.; Smrž, M.; Volodin, V. A.; Mocek, T.; Macak, J. M.; Bulgakova, N. M. Non-thermal regimes of laser annealing of semiconductor nanostructures: crystallization without melting. *Front. Nanotechnol*. **2023**, *5*, 1271832.
9. Volodin, V. A.; Cheng, Y.; Bulgakov, A. V.; Levy, Y.; Beranek, J.; Nagisetty, S. S.; Zukerstein, M.; Popov, A. A.; Bulgakova, N. M. Single-shot selective femtosecond and picosecond infrared laser crystallization of an amorphous Ge/Si multilayer stack *Opt. Laser Technol*. **2023**, *161*, 109161.
10. Chichkov, B. N.; Momma, C.; Nolte, S.; von Alvensleben, F.; Tünnermann, A. Femtosecond, picosecond and nanosecond laser ablation of solids. *Appl. Phys. A* **1996**, *63*, 109–115.
11. Pey K. L., Lee, P.S. Pulsed laser annealing technology for nanoscale fabrication of silicon-based devices in semiconductors. In: *Advances in Laser Materials Processing: Technology, Research and Applications*; Lawerence, J.; Pou, J.; Low, D.K.Y.; Toyserkani, E., Eds.; Woodhead Publishing Limited, 2010, pp. 327-364. https://doi.org/10.1533/9781845699819.4.327
12. Sameshima, T.; Usui, S.; Pulsed laser-induced amorphization of silicon films. *J. Appl. Phys*. **1991**, *70*, 1281-1289.
13. Dassow, R.; Köhler, J. R.; Helen, Y.; Mourgues, K.; Bonnaud, O.; Mohammed-Brahim T.; Werner, J. H. Laser crystallization of silicon for high-performance thin-film transistors. *Semicond. Sci. Technol*. **2000**, *15*(10), L31-L34.
14. Zhou, R. J.; Zheng, Y.; Qian, L.; Yang, Y. X.; Holloway, P. H.; Xue, J. G. Solution-processed, nanostructured hybrid solar cells with broad spectral sensitivity and stability. *Nanoscale* **2012**, *4*, 3507-3514.
15. Stuchlikova, T. H.; Remes, Z.; Stuchlik, J. Germanium and tin nanoparticles encapsulated in amorphous silicon matrix for optoelectronic applications. In Proceedings of 10th Anniversary International Conference on Nanomaterials – Research & Applications (Nanocon-2018), Brno, Czech Republic, October 17-19, 2018; p. 226-229.
16. Akl A. A.; Howari, H. Nanocrystalline formation and optical properties of germanium thin films prepared by physical vapor deposition. *J. Phys. Chem. Sol*. **2009**, *70*, 1337–1343.
17. Zhao, Z. X.; Sun, M. H.; Yang, C.; Zhang, C. L. Electrodeposition of crystalline germanium thin films by the electrochemical liquid-liquid-solid method, *J. Electroanal. Chem*. **2023**, *948*, 117829.
18. Liu, J.; Camacho-Aguilera, R.; Bessette, J.T.; Sun, X.; Wang, X.; Cai Y.; Y., Kimerling, L. C.; Michel, J. Ge-on-Si optoelectronics. *Thin Solid Films* **2012**, *520*, 3354–3360.
19. Fischer, I.A.; Brehm, M., De Seta, M., Isella, G., Paul, D.J., Virgilio, M., Capellini, G. On-chip infrared photonics with Si-Ge-heterostructures: What is next? *APL Photonics* **2022**, *7*, 050901.

20. Liu, Z.; Zhou, T.; Li, L.; Zuo, Y.; He, C.; Li, C.; Xue, C.; Cheng, B.; Wang, Q. Ge/Si quantum dots thin film solar cells. *Appl. Phys. Lett*. **2013**, *103*, 082101.
21. Li, C.; Ni, J.; Sun, X.; Wang, X.; Li, Z,; Cai, H.; Li, J.; Zhang, J. Nanocrystalline germanium nip solar cells with spectral sensitivities extending into 1450 nm. *J. Phys. D: Appl. Phys*. **2017**, *50*, 045108.
22. Pchelyakov, O.P.; Bolkhovityanov, Y.B.; Dvurechenskiĭ, A.V.; Sokolov, L.V.; Nikiforov, A.I.; Yakimov, A.I.; Voigtländer, B. Silicon-germanium nanostructures with quantum dots: Formation mechanisms and electrical properties. *Semiconductors* **2000**, *34*, 1229-1247.
23. Yang, L.; Watling, J. R.; Wilkins, R. C. W.; M. Boriçi, M.; Barker, J. R.; Asenov, A.; Roy, S. Si/SiGe heterostructure parameters for device simulations. *Semicond. Sci. Technol*. **2004**, *19*, 1174-1182.
24. Cheng, Y.; Bulgakov, A.V.; Bulgakova, N.M.; Beránek, J.; Milekhin, I.A.; Popov, A.A.; Volodin, V.A. Ultrafast infrared laser crystallization of amorphous Ge films on glass substrates. *Micromachines* **2023**, *14*, 2048.
25. Liu, J.M. Simple technique for measurements of pulsed Gaussian-beam spot sizes. *Opt. Lett*. **1982**, *7*, 196–198.
26. Clark, A.H. Electrical and optical properties of amorphous germanium. *Phys. Rev*. **1967**, *154*, 750-757.
27. Liu, J.M. Simple technique for measurements of pulsed Gaussian-beam spot sizes. *Opt. Lett.* **1982**, *7*, 196-198. https://doi.org/10.1364/OL.7.000196
28. Demichelis, F.; Minetti-Mezzetti, E.; Tagliaferro, A.; Tresso, E.; Rava, P.; Ravindra, N.M. Optical properties of hydrogenated amorphous silicon. *J. Appl. Phys*. **1986**, *59*, 611-618. https://doi.org/10.1063/1.336620
29. Smith, J.E., Jr.; Brodsky, M.H.; Crowder, B.L.; Nathan, M.I.; Pinczuk, A. Raman spectra of amorphous Si and related tetrahedrally bounded semiconductors. *Phys. Rev. Lett.* **1971**, *26*, 642-646.
30. Shen, S.C.; Fang, C.J.; Cardona, M.; Genzel, L. Far-infrared absorption in pure and hydrogenated *a*-Ge and *a*-Si. *Phys. Rev. B.* **1980**, *22*, 2913-2919.
31. Bermejo, D.; Cardona, M. Raman scattering in pure and hydrogenated amorphous germanium and silicon. *J. Non-Cryst. Solids*. **1979**, *32*, 405-419. https://doi.org/10.1016/0022-3093(79)90085-1
32. Volodin, V. A.; Efremov, M. D.; Deryabin, A.S.; Sokolov, L.V. Determination of the composition and stresses in $Ge_xSi_{(1-x)}$ heterostructures from Raman spectroscopy data: Refinement of model parameters. *Semiconductors* **2006**, *40*, 1314-1320.
33. Sugawa, S.; Yokogawa, R.; Yoshioka, K.; Arai, Y.; Yonenaga, I.; Ogura, A. Temperature dependence of Raman peak shift in single crystalline silicon-germanium. *ECS J. Solid State Sci. Technol.* **2023**, *12*, 064004. https://doi.org/10.1149/2162-8777/acdffa
34. Renucci, M.A.; Renucci, J.B.; Cardona, M. Raman scattering in Ge-Si alloys. In *Proceedings of the Second International Conference on Light Scattering in Solids*, Flammarion, Paris (1971), pp. 326-329.
35. Mooney, P.M.; Dacol, F.H.; Tsang, J.C.; Chu, J. O. Raman scattering analysis of relaxed $Ge_xSi_{1-x}$ alloy layers. *Appl. Phys. Lett*. **1993**, *62*, 2069–2071. https://doi.org/10.1063/1.109481
36. Grevtsov, N., Chubenko, E.; Bondarenko, V.; Gavrilin, I.; Dronov, A.; Gavrilov, S.; Goroshko, D.; Goroshko, O.; Rymski, G; Yanushkevich, K. Thermoelectric materials based on cobalt-containing sintered silicon-germanium alloys. *Mater. Res. Bull*. **2025**, *284*, 113258. https://doi.org/10.1016/j.materresbull.2024.113258
37. Wojdyr, M. Fityk: A General-Purpose Peak Fitting Program. *J. Appl. Cryst*. **2010**, *43*, 1126-1128. https://doi.org/10.1107/S0021889810030499
38. Zhigunov, D.M.; Kamaev, G.N.; Kashkarov, P.K.; Volodin, V.A. On Raman scattering cross section ratio of crystalline and microcrystalline to amorphous silicon. *Appl. Phys. Lett*. **2018**, *113*, 023101. https://doi.org/10.1063/1.5037008
39. Hao, Z.; Kochubei, S.A.; Popov, A.A.; Volodin V.A. On Raman scattering cross section ratio of amorphous to nanocrystalline germanium. *Solid State Commun*. **2020**, *313*, 113897. https://doi.org/10.1016/j.ssc.2020.113897
40. Volodin, V.A.; Sachkov, V.A. Improved model of optical phonon confinement in silicon nanocrystals. *J. Exp. Theor. Phys*. **2013**, *116*, 87–94. https://doi.org/10.1134/S1063776112130183
41. Khoo, K.H.; Zayak, A.T.; Kwak, H.; Chelikowsky, J.R. First-principles study of confinement effects on the Raman spectra of Si nanocrystals. *Phys. Rev. Lett*. **2010**, *105*, 115504. https://doi.org/10.1103/PhysRevLett.105.115504

42. Jie, Y.; Wee, A.T.S.; Huan, C.H.A.; Shen, Z.X.; Choi, W.K. Phonon confinement in Ge nanocrystals in silicon oxide matrix. *J. Appl. Phys.* **2011**, *109*, 033107. https://doi.org/10.1063/1.3503444
43. Zi, J.; Zhang, K.; Xie, X. Comparison of models for Raman spectra of Si nanocrystals. *Phys. Rev. B* **1997**, *55*, 9263-9266. https://doi.org/10.1103/PhysRevB.55.9263
44. Jang, B.K.; Yoshiya, M.; Matsubara, H. Influence of number of layers on thermal properties of nano-structured zirconia film fabricated by EB-PVD method. *J. Japan Inst. Met. Mater*. **2005**, *69*, 56-60. https://doi.org/10.2320/jinstmet.69.56